\documentclass[12pt]{iopart}

\usepackage{iopams}
\usepackage{graphicx}
\usepackage{indentfirst}    

\begin{document}

\title[Colliding ionization injection in a beam driven plasma accelerator]{Colliding ionization injection in a beam driven plasma accelerator}

\author{Y. Wan$^{1,3}$, C. J. Zhang$^{1,3}$, F. Li$^{1,3}$, Y. P. Wu$^1$, J. F. Hua$^1$, C.-H. Pai$^1$, W. Lu$^{1}$, C. Joshi$^2$, W. B. Mori$^2$, and Y. Q. Gu$^3$}
\address{$^1$ Department of Engineering Physics, Tsinghua University, Beijing 100084, China}
\address{$^2$ University of California Los Angeles, Los Angeles, CA 90095, USA}
\address{$^3$ Laser Fusion Research Center, China Academy of Engineering Physics, Mianyang, Sichuan 621900, China}

\ead{weilu@tsinghua.edu.cn, jfhua@tsinghua.edu.cn,chpai@tsinghua.edu.cn}
\begin{abstract}
The proposal of generating high quality electron bunches via ionization injection triggered by an counter propagating laser pulse inside a beam driven plasma wake is examined via two-dimensional particle-in-cell simulations. It is shown that electron bunches obtained using this technique can have extremely small slice energy spread, because each slice is mainly composed of electrons ionized at the same time. Another remarkable advantage is that the injection distance is changeable. A bunch with normalized emittance of 3.3 nm, slice energy spread of 15 keV and brightness of $7.2\times 10^{18}$ A m$^{-2}$ rad$^{-2}$ is obtained with an optimal injection length which is achieved  by adjusting the launch time of the drive beam or by changing the laser focal position. This makes the scheme a promising approach to generate high quality electron bunches for the fifth generation light source.
\end{abstract}

\maketitle

\section{Introduction}
Acceleration of electrons via plasma based wake accelerators has attracted much attention since the past decades\cite{esarey2009}. The longitudinal electric fields of such wakes are orders of magnitude higher than the conventional cavity-based accelerators ($\sim$ 100 MV/m). It has been demonstrated that electrons can be accelerated up to GeV energies in centimeters nonlinear plasma wakes induced by a powerful laser pulse\cite{leemans2006,kneip2009,froula2009,clayton2010,wang2013} and more than 42 GeV in a meter-scale plasma using a high charge electron beam driver\cite{muggli2004,hogan2005,blumenfeld2007,litos2014}. In the latter case (beam driver)\cite{lu2006,lu2006pop,rosenzweig1991}, when a dense, ultrarelativistic electron beam propagates through a plasma, the plasma electrons will be blown out completely by the beam's coulomb force leaving behind the pure ions cavity which will then pull the electrons back, creating a wakefield with a phase velocity equal to the beam's velocity. This wake produces high longitudinal electric field ideal for acceleration, as well as quasi-linear focusing force on the accelerated electrons due to ion cavities.

In the plasma based accelerators, the controllable injection of electrons into the wakefield bucket is of paramount importance, since it affects the beam quality (charge, energy spread, emittance, current, and brightness) dramatically. Self-injection is quite simple, but it is not controllable\cite{plateau2012,wiggins2010}. Besides, various injection techniques have been developed and demonstrated, such as ponderomotive force injection\cite{umstadter1996}, injection via external magnetic field\cite{vieira2011}, colliding pulse injection\cite{esarey1997,faure2006} and plasma density transition injection\cite{suk2001,geddes2008}. Another promising scheme is ionization injection\cite{clayton2010,oz2007,rowlands2008,pak2010,mcguffey2010,liu2011,pollock2011} where electrons are produced inside the bucket by the electric field of a laser pulse or an electron beam driver, so that they can be captured and accelerated much easier. Recently, it was proposed to combine the ionization injection with a co-propagating laser pulse\cite{hidding2012}. This approach can further reduce the injected electrons's transverse emittance to several tens of nm by using a low intensity ionizing laser, which is attractive for x-ray free electron laser sources\cite{emma2010}. However, it has drawbacks of large slice energy spread due to the long injection distance.\cite{xu2014}. Another big try is to use two transverse colliding laser pulses as injection triggers\cite{lifei2013}. The ionization only occurs in the region where two counter propagating lasers overlap with each other, so that the injection distance is greatly reduced, leading to small slice energies. However, this scheme is quite complex to realize in experiment.

In this paper, we propose a simple and effective ionization injection scheme. The injected electrons into a beam-driven plasma wakefield accelerators are ionized via tunnel ionization at the focus of a counter propagating laser pulse (in a beam driven plasma wake). This scheme is simple and very similar to the layout of ordinary Thomson Scattering experiment, which makes it easier to realize. Electron beams generated in this scheme have extremely small slice energy spread ($\sim15$ keV), and their total emittance are also reduced to $3\sim5$ nm by changing the launching time of beam driver or laser injector or its focal position to optimize the injection distance, which make this scheme attractive in generating high quality electron beams for free electron laser.

\section{Theoretical analysis and simulations}
The mechanism for colliding ionization injection is explored using the 2D Particle-in-Cell (PIC) code OSIRIS\cite{fonseca2002} in Cartesian coordinates using a fixed window. We define the z axis as the drive beam's propagating direction, and the x axis as the transverse direction. The simulation window is $610\times101$ $\mu$m, which is divided into $14400\times3200$ cells along the z and x direction respectively. ADK (Ammosov-Delone-Krainov) model is employed as the tunneling ionization model in the code\cite{ammosov1986}.

The plasma consists of pre-ionized part with the density of $n_p=2.4\times10^{17}$ cm$^{-3}$ and neutral He atoms with density of $n_{He}=1.34\times10^{18}$ cm$^{-3}$. 4 particles for electron and 8 particles for Helium are initialized in each cell. A 100 MeV electron beam with the transverse and longitudinal dimensions $\sigma_r=3.8\ \mu$m, $\sigma_z=6\ \mu$m, and $n_b=[N/(2\pi)^{3/2}\sigma_z\sigma_r^2]\exp(-r^2/2\sigma_r^2-z^2/2\sigma_z^2)
=5.1\times10^{17}$ cm$^{-3}$ propagates through the plasma and excites a blowout wakefield with the wavelength of $\lambda_p\approx80\ \mu$m. The beam's self-electric field ($\sim$ 30 GV/m) does not ionize the helium atoms. Meanwhile, a counter propagating laser pulse is synchronized with the electron beam. The laser has a normalized vector potential $a_0=0.03$, a pulse duration $\tau=20 $fs, and a focal spot size $w_0=2\ \mu$m. These parameters correspond to a focused intensity of $2\times10^{15}$ W/cm$^2$ for $\lambda_0=800$ nm.

As the laser's propagation direction is opposite to the driver, the ionization phase region of the wakes is quite long, about $2R_0$, where $R_0$ is the physical ionization region of the laser, the same order as two times of the Rayleigh length of the laser pulse,$2z_R=2\pi w_0^2/\lambda_0$ at most times. In our case, $R_0\approx55\ \mu$m.
It is very different from the co-propagating injection case where ionization fills only a small phase region. In addition, the trapping threshold\cite{xu2014,pak2010} is given by $\Delta\psi\approx\psi-\psi_{init}<-1$ where $\psi=e(\phi-A_z)/mc^2$ is the normalized wake potential and $\psi_{init}$ is the wake potential where an electron is released. So there is also a phase region that satisfies this trapping condition, which covers about $45\ \mu$m in our simulation.

\begin{figure}[htbp]
  \centering
  \includegraphics[width=0.5\textwidth]{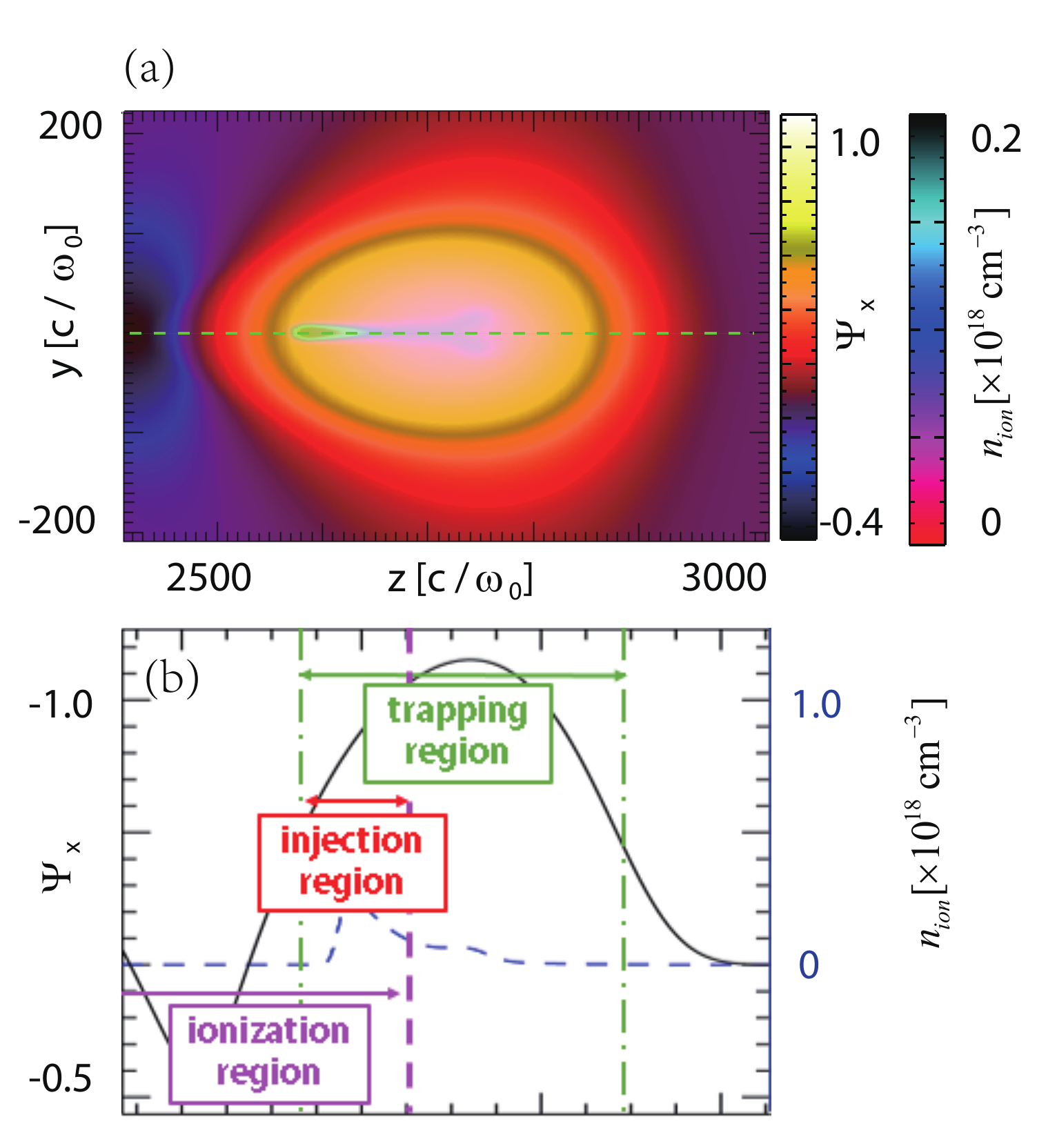}\\
  \caption{(a) The normalized wake potential $\psi$ distribution and ionization level at he beginning of collision by the laser pulse. (b) The lineout on the dashed line in (a). Laser-ionized He$^{+}$ are shown with dashed line and the wake potential with solid line. The purple, green and red ares represent the actually ionization phase region, the trapping phase region and the injection phase regions, respectively.}\label{psi}
\end{figure}

Then by changing the launching time of beam driver or laser injector or its focal position to adjust the overlap of trapping and ionization phase regions, it is possible to ensure the injection distance short enough and thus to obtain an electron bunch with extremely small transverse emittance.

As shown in Fig.~\ref{psi} (a), the ionization is set to occur from the start position of the acceleration phase. ($\partial\psi/\partial\xi<0$). In Fig.~\ref{psi} (b), a lineout corresponding to the green dashed line of Fig.~\ref{psi}(a) is presented. It is shown that $\psi_{min}\approx-$ 0.5, so electrons ionized at $\xi_i$ that satisfies $\psi(\xi)\geq 0.5$ can be trapped, which is represented by the green region in this inset. And the purple region corresponds to the ionization region. There is a small overlap region (the red area in Fig.~\ref{psi} (b)) between ionization (the blue area)and trapping region (the purple area) where ionized electrons can be finally trapped by the wake and then be accelerated. In this way, the injection distance can be limited to a short range and the final emittance will be reduced greatly. The amount of injected charge is proportional to the neutral Helium density.

\begin{figure}[htbp]
  \centering
  \includegraphics[width=0.8\textwidth]{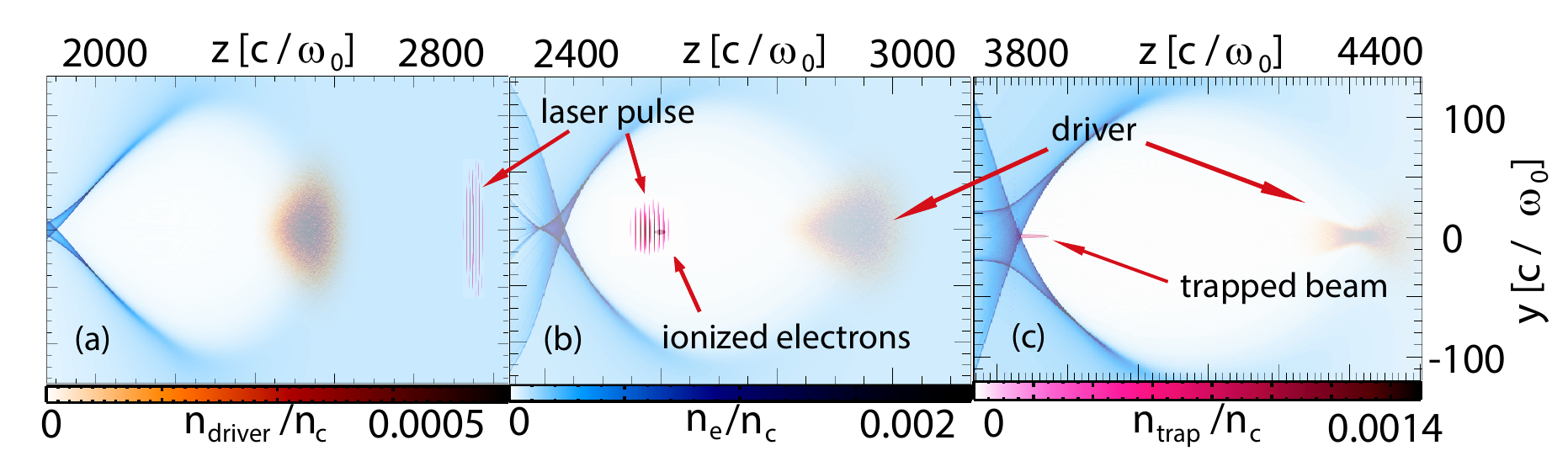}\\
  \caption{Snapshots of (a)-(c) show the charge density distribution of driver beam, wake electrons and helium electrons at three different times. (a) $\sim$160 fs before laser pulse ionizes, (b) at he initial time of ionization, (c) $\sim$ 600 fs after the injected electrons become trapped in the wake.}\label{injection}
\end{figure}

Fig.~\ref{injection} illustrate the injection process in detail, where $n_c=1.74\times10^{21}$ cm$^{-3}$ is the plasma critical density for 800 nm laser pulse. In Fig.~\ref{injection} (a), the laser is moving backward and in Fig.~\ref{injection} (b), it begins to ionized electrons located near the threshold of trapping condition, not yet having reached its focal point at $z=2360\ c/\omega_0$, where $\omega_0 $ is the laser pulse's frequency. Then these electrons respond to the wake fields and are rapidly accelerated to a longitudinal velocity close to c as they slip backwards to the back of the bucket. They are finally trapped and start to be accelerated by the wake, as depicted in Fig.~\ref{injection} (c). Meanwhile, along the propagation direction of the laser, the rest of ionized electrons either become the wake background or are perhaps trapped by the following wake, which don't affect the injected beam's quality.

\begin{figure}[htbp]
  \centering
  \includegraphics[width=0.8\textwidth]{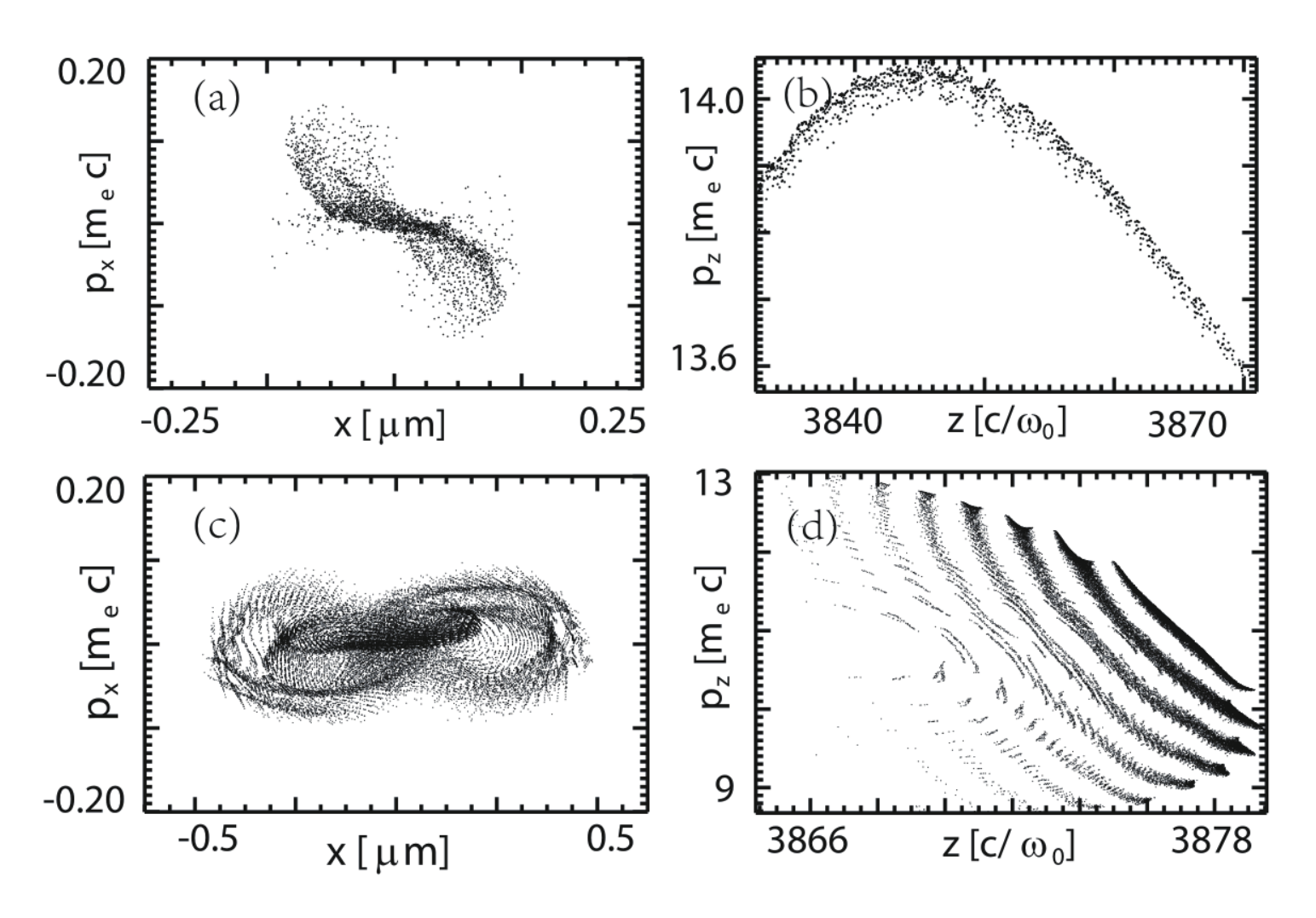}\\
  \caption{Comparison of The (a) (c) $x-p_x$, (b) (d) $z-p_z$ phase space distribution about 600 fs after ionization between colliding ionization injection (a),(b) and the co-propagation injection (c),(d). }\label{phase_space}
\end{figure}

Fig.~\ref{phase_space} (a) and (b) show the transverse phase space ($x-p_x$) and longitudinal phase space ($z-p_z$) of injected electrons about 600 fs after ionization, respectively. The normalized emittance $\epsilon_n=\sqrt{\langle x^2\rangle\langle p_x^2\rangle-\langle xp_x\rangle^2}$ is calculated. The injected beam has an ultralow projected transverse $\epsilon_n$ for the whole bunch of about 3.3 nm rad, after propagating about 180 $\mu$m. At this point, the beam has an average energy of 6.67 MeV, with a rms energy spread of 80 keV, a slice rms energy spread (about $0.5\ \mu$m thickness) of $\sim$13 keV, and a total charge of about 0.4 pC. The energy chirp is reduced via beam loading effect. The beam has a pulse duration of 4.5 fs and peak current $I_p\approx$ 40 A. The brightness $B_n\approx 2I_p/\epsilon_n^2\approx 7.2\times 10^{18}$ A m$^{-2}$rad$^{-2}$, is about 2 or 3 orders of magnitude larger than that in the Linac Coherent Light Source (LCLS)\cite{emma2010}. The total charge can be increased by launching an injector laser with a larger focal spot size and the energy spread can also be further reduced by optimizing beam loading or by using a longer wavelength wake.

As single pulse longitudinal injection schemes are easier to realize in experiments than multi pulses injection, here we only compare this scheme with the co-propagating injection case.

To obtain ultralow transverse emittance $\epsilon_n$ in the wakefield, , several key factors have to be considered. First, the initial thermal emittance $\epsilon_{th}=\sigma_{p_{x0}}\sigma_{x0}$ should be as small as possible. In the beam driven plasma wakes, the beam's self electric field is too low to ionize the doping gas (e.g.Helium), while an injector laser of $800$ nm with much lower intensity (e.g.$a_0=0.03$) and very small focal spot size can be used to trigger further ionization to obtain beams with much smaller initial momentum and beam size.

The second important factor is the injection distance\cite{wang2002,xu2014}. It can lead to the phase mixing process, which occurs when electrons are born at different time and therefore at different phases of their betatron oscillations, further leading to the increase of emittance. In the colliding ionization injection case, the injection distance can be reduced by adjusting the launching time of the driver and injector or the focal position of injector. In the simulation above, it is only about 1/4 of the whole ionization distance. However, in the co-propagating injection scheme, the ionization happens approximately at some fixed phase (usually at the position where $\psi$ is maximum), and thus the injection distance is generally fixed and nearly the same as the ionization distance. So the phase mixing process is more severe.

Another vital factor is the energy spread, as electrons with different energies are rotating with different betatron frequencies, also leading to phase mixing. The energy spread of injected electrons are mainly from three aspects: initial ionized momentum spread (proportional to $a_0$), the longitudinal acceleration phase spread (spread in $E_z$) and slice energy spread. In most cases, the longitudinal $\delta E_z$ can be reduced by beam loading. Meanwhile, slice energy spread results from the injection process itself, which is not easy to decrease.

It is known that $\psi$ in the ion cavity can be expressed as $\psi(\xi,r)\approx[r_b^2(\xi)-r^2]/4$, where $r_b(\xi)$ is the normalized radius of the blowout at different $\xi$ and $r_b^2(\xi)=r_m^2-\xi^2$, where $r_m$ is the largest blowout radius\cite{lu2006,lu2006pop,xu2014}. By applying the trapping condition $\delta\psi\approx-1$, it is straightforward to obtain:
\begin{eqnarray}
\centering
\xi_f^2+r_f^2=4+\xi_i^2+r_i^2 \label{lonrelation1}
\end{eqnarray}
where f and i are the final and initial positions, respectively. In most cases, especially when the injector laser's spot size is much smaller than the blowout radius, $r_f$ and $r_i$ can be neglected. So Eq.~\ref{lonrelation1} can be simplified as:
\begin{eqnarray}
\centering
\xi_f=\sqrt{4+\xi_i^2} \label{lonrelation2}
\end{eqnarray}
In the co-propagating injection case, $\xi_i$ is generally unchangeable during the ionization process. Therefore, each final slice of the injected beam is composed of electrons ionized at different time. This leads to energy difference for electrons within a single slice. However, in the colliding injection case, $\xi_i$ is varying during the ionization process and thus each final slice is only related to electrons born at the same time. As a result, the injected electrons have extremely small slice energy spread as well as slice emittance.

\begin{figure}[htbp]
  \centering
  \includegraphics[width=0.8\textwidth]{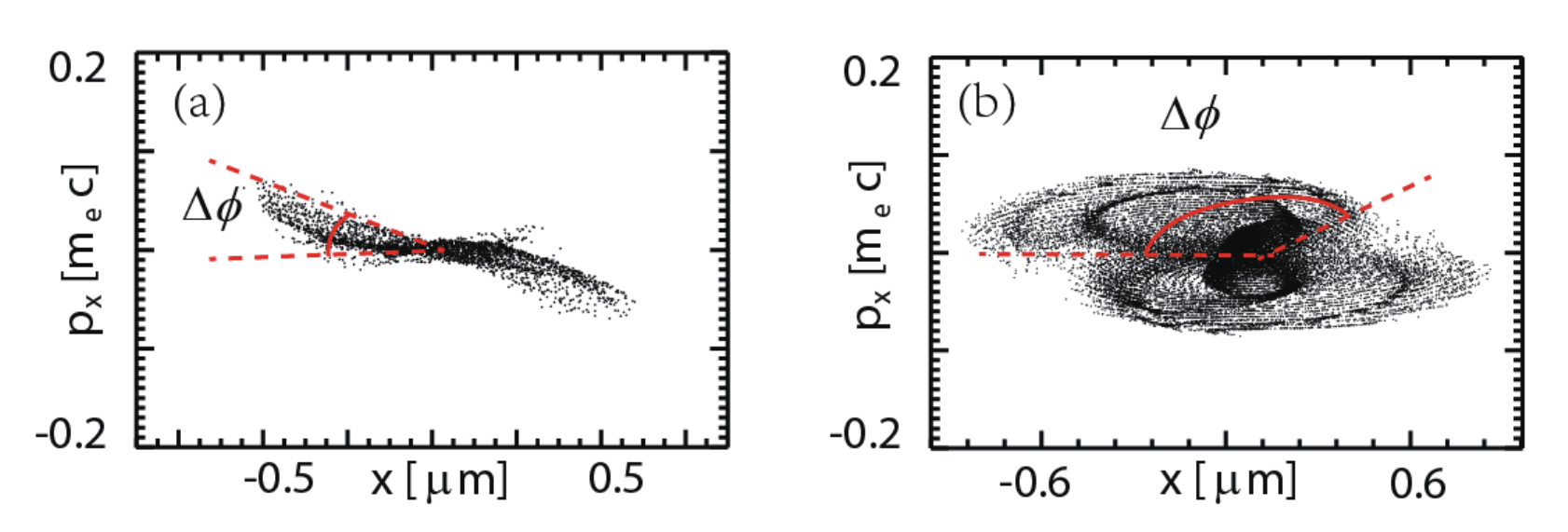}\\
  \caption{Comparison of $x-p_x$ phase space distribution about 130fs fs after ionization between the colliding ionization injection (a) and the co-propagation injection (b).}\label{initial_phase_space}
\end{figure}

To verify the analysis above in detail, we have also simulated the co-propagating injection scheme. In this case, an injection laser pulse propagates collinearly at an optimum distance (about 78 fs) behind the beam driver, where $E_z=0$. The drive beam, injection laser and pre-ionized plasma are the same as the simulation for the colliding ionization injection  described above. The neutral helium density is $3.1\times10^{16}$ cm$^{-3}$, which is different from the colliding propagation case, for the purpose of similar injected charge( about 0.45 pC). Fig.~\ref{phase_space} (c) and (d) show the phase space of injected electrons of different planes about 560 fs after ionization via the co-propagating approach. It is shown that the phase mixing in the co-propagating case is much more severe than the colliding scheme and its final emittance is about $10.5$ nm rad. Fig.~\ref{phase_space} (d) shows that the electron's slice energy is also very large, about $0.4$ MeV and their total energy spread is about 0.5 MeV. The beam has a pulse duration of 1 fs. The layer-like structure in this $z-p_z$ phase space is due to the initial electrons distribution after ionization by the laser pulse. Only electrons located near the peaks of the laser electric field are ionized, so initially the electrons have a longitudinal periodic structure. Fig.~\ref{initial_phase_space} further compares the injection distances of these two schemes, taken at about 130 fs after the onset of injection. As seen in Fig.~\ref{initial_phase_space}(b), because the injection is ongoing, the first ionized electrons have rotated around $3/4\pi$, while the final ionized electrons have just been released. However, in our proposed scheme, the injection have already finished, and the phase difference $\Delta\phi$ is much smaller than the co-propagating case.

\begin{figure}[htbp]
  \centering
  \includegraphics[width=0.9\textwidth]{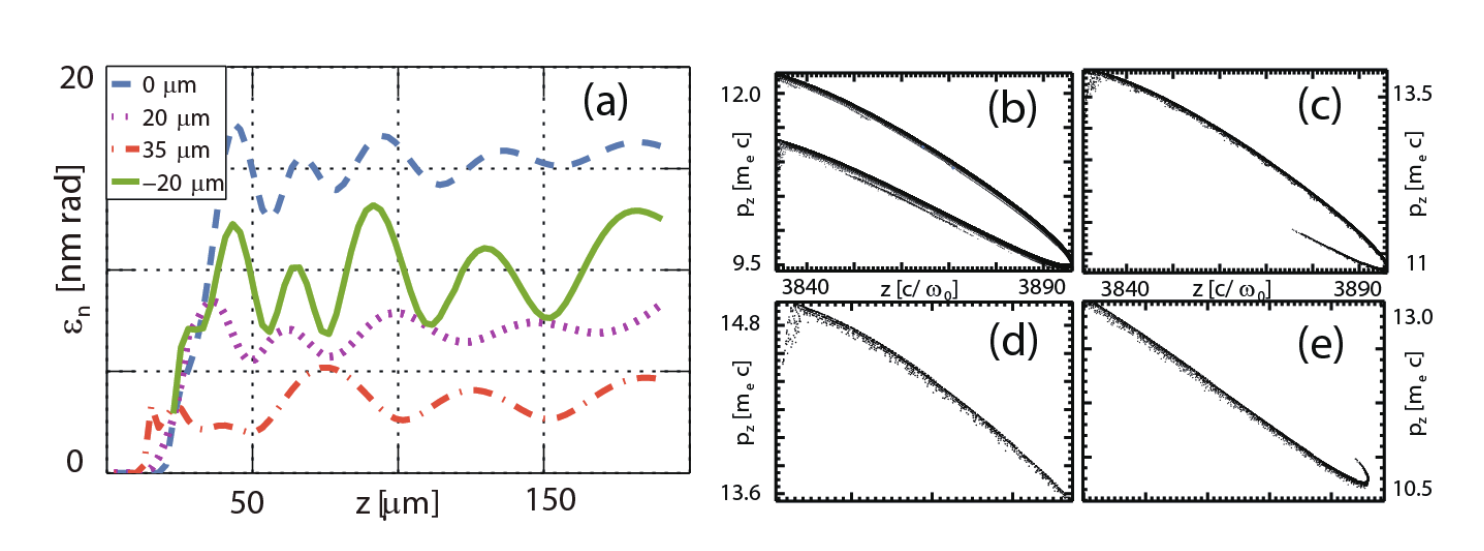}\\
  \caption{The impact of focal position on $\epsilon_n$. Variation of $\epsilon_n$ evolution (a) and $z-p_z$ phase space distribution (b)(c)(d)(e) under different focal positions. The (b), (c), (d) and (e) correspond with the focal position $z_f$=0, 20, 35, and $-$ 20 $\mu$m.  }\label{focal_position_compare}
\end{figure}

To obtain the extremely low transverse emittance in the colliding injection scheme, the key point is to adjust the launching time of beam driver or the injector laser's focal position to optimize the injection length. Here we study the effect of this issue by changing the focal position $z_f$. We define the $z_f=0$ as the laser focuses at $E_z=0$ when it encounters with the $\psi_{max}$ of the wake, and $z_f<0$ as the focal position at $E_z>0$ (deceleration phase). The simulation parameters are identical to those of Fig.~\ref{phase_space} (a), except that the neutral helium density is set as $5\times10^{16}$cm$^{-3}$. Four cases of $z_f$=0, 20, 35, and -20$\mu$ m are simulated. As $z_f$ increase, the overlap phase area of ionization distance and trapping phase region is generally reduced, and hence the injection distance also become shorter, leading to the reduction of final total emittance, which is consistent to the fact that the emittance of $z_f=35\mu$m is the smallest as shown in Fig.~\ref{focal_position_compare} (a). Through more simulation tests, it is found that the final emittance can be less than 7nm rad as long as the focal position satisfies that 20 $\mu$m$\leq z_f\leq$50 $\mu$m.

Fig.~\ref{focal_position_compare} (b) to (e) illustrate the longitudinal phase space of $z_f$=0, 20, 35, and -20 $\mu$m, respectively. If the laser pulse converges upon the point $E_z=0$, electrons at both $\xi_i$ and $-\xi_i$ are ionized. According to Eq.~\ref{lonrelation2}, they will finally be trapped at the same acceleration phase $\xi_f$. However, they will experience different injection distances. Electrons born at $-\xi_i$ where $E_z>0$ will first be accelerated backward and then be decelerated when they enter the phase of $E_z<0$ and finally arrives at $\xi_i$ with initial momentum. So obviously, electrons born at $\xi_i$ arrive at the final trapped phase $\xi_f$ earlier than those born at $-\xi_i$. Therefore, in Fig.~\ref{focal_position_compare} (b), (c) and (e), some slices have double energies. Also because of shorter injection distance of electrons born at $\xi_i>0$, the emittance of $z_f = 20\ \mu$m is relative smaller than that of $z_f = -20\ \mu$m, as shown in Fig.\ref{focal_position_compare}(a).


\section{Conclusions}
A new ionization injection scheme is proposed in the context of beam driven plasma accelerators. Electrons are injected via a counter propagating laser pulse. The platform for this approach is very similar to that of the Thomson Scattering experiments, and the overlaps of both time and space between the injector laser pulse and electron driver from conventional accelerators are much easier to perform than the co-propagating injection scheme in experiments. Besides, electron bunches obtained from this scheme have extremely small slice energy spread($\sim$ 15 keV). Compared with the co-propagating case, another remarkable advantage is that the injection distance is changeable. By optimizing the injection distance, the total emittance of the injected bunch can be reduced to $3 \sim 5$ nm, making this novel scheme a promising approach to produce high quality electron beams for the next generation light sources.

\section{Acknowledgement}
This work is supported by National Natural Science Foundation of China
No.11375006, No. 11425521, No. 11535006, No. 11475101 and No.11175102.

\section*{References}
\bibliographystyle{unsrt}
\bibliography{ref}

\begin{thebibliography}{10}

\bibitem{esarey2009}
E.~Esarey, C.~B. Schroeder, and W.~P. Leemans.
\newblock {\em Reviews of Modern Physics}, 81(3):1229, 2009.

\bibitem{leemans2006}
W.~P. Leemans, B.~Nagler, A.~J. Gonsalves, Cs~Toth, K.~Nakamura, C.~G.~R.
  Geddes, E.~Esarey, C.~B. Schroeder, and S.~M. Hooker.
\newblock {\em Nature Physics}, 2(10):418, 2006.

\bibitem{kneip2009}
S.~Kneip, S.~R. Nagel, S.~F. Martins, S.~P.~D. Mangles, C.~Bellei, O.~Chekhlov,
  R.~J. Clarke, N.~Delerue, E.~J. Divall, G.~Doucas, K.~Ertel, F.~Fiuza,
  R.~Fonseca, P.~Foster, S.~J. Hawkes, C.~J. Hooker, K.~Krushelnick, W.~B.
  Mori, C.~A.~J. Palmer, K.~Ta Phuoc, P.~P. Rajeev, J.~Schreiber, M.~J.~V.
  Streeter, D.~Urner, J.~Vieira, L.~O. Silva, and Z.~Najmudin.
\newblock {\em Physical Review Letters}, 103(3):035002, 2009.

\bibitem{froula2009}
D.~H. Froula, C.~E. Clayton, T.~Doeppner, K.~A. Marsh, C.~P.~J. Barty,
  L.~Divol, R.~A. Fonseca, S.~H. Glenzer, C.~Joshi, W.~Lu, S.~F. Martins,
  P.~Michel, W.~B. Mori, J.~P. Palastro, B.~B. Pollock, A.~Pak, J.~E. Ralph,
  J.~S. Ross, C.~W. Siders, L.~O. Silva, and T.~Wang.
\newblock {\em Physical Review Letters}, 103(21):215006, 2009.

\bibitem{clayton2010}
C.~E. Clayton, J.~E. Ralph, F.~Albert, R.~A. Fonseca, S.~H. Glenzer, C.~Joshi,
  W.~Lu, K.~A. Marsh, S.~F. Martins, W.~B. Mori, A.~Pak, F.~S. Tsung, B.~B.
  Pollock, J.~S. Ross, L.~O. Silva, and D.~H. Froula.
\newblock {\em Physical Review Letters}, 105(10):105003, 2010.

\bibitem{wang2013}
Xiaoming Wang, Rafal Zgadzaj, Neil Fazel, Zhengyan Li, S.~A. Yi, Xi~Zhang,
  Watson Henderson, Y.~Y. Chang, R.~Korzekwa, H.~E. Tsai, C.~H. Pai,
  H.~Quevedo, G.~Dyer, E.~Gaul, M.~Martinez, A.~C. Bernstein, T.~Borger,
  M.~Spinks, M.~Donovan, V.~Khudik, G.~Shvets, T.~Ditmire, and M.~C. Downer.
\newblock {\em Nature Communications}, 4:2988, 2013.

\bibitem{muggli2004}
P.~Muggli, B.~E. Blue, C.~E. Clayton, S.~Deng, F.~J. Decker, M.~J. Hogan,
  C.~Huang, R.~Iverson, C.~Joshi, T.~C. Katsouleas, S.~Lee, W.~Lu, K.~A. Marsh,
  W.~B. Mori, C.~L. O'Connell, P.~Raimondi, R.~Siemann, and D.~Walz.
\newblock {\em Physical Review Letters}, 93(1):014802, 2004.

\bibitem{hogan2005}
M.~J. Hogan, C.~D. Barnes, C.~E. Clayton, F.~J. Decker, S.~Deng, P.~Emma,
  C.~Huang, R.~H. Iverson, D.~K. Johnson, C.~Joshi, T.~Katsouleas, P.~Krejcik,
  W.~Lu, K.~A. Marsh, W.~B. Mori, P.~Muggli, C.~L. O'Connell, E.~Oz, R.~H.
  Siemann, and D.~Walz.
\newblock {\em Physical Review Letters}, 95(5):054802, 2005.

\bibitem{blumenfeld2007}
Ian Blumenfeld, Christopher~E. Clayton, Franz-Josef Decker, Mark~J. Hogan,
  Chengkun Huang, Rasmus Ischebeck, Richard Iverson, Chandrashekhar Joshi,
  Thomas Katsouleas, Neil Kirby, Wei Lu, Kenneth~A. Marsh, Warren~B. Mori,
  Patric Muggli, Erdem Oz, Robert~H. Siemann, Dieter Walz, and Miaomiao Zhou.
\newblock {\em Nature}, 445(7129):nature05538, 2007.

\bibitem{litos2014}
M.~Litos, E.~Adli, W.~An, C.~I. Clarke, C.~E. Clayton, S.~Corde, J.~P.
  Delahaye, R.~J. England, A.~S. Fisher, J.~Frederico, S.~Gessner, S.~Z. Green,
  M.~J. Hogan, C.~Joshi, W.~Lu, K.~A. Marsh, W.~B. Mori, P.~Muggli,
  N.~Vafaei-Najafabadi, D.~Walz, G.~White, Z.~Wu, V.~Yakimenko, and G.~Yocky.
\newblock {\em Nature}, 515(7525):13882, 2014.

\bibitem{lu2006}
W.~Lu, C.~Huang, M.~Zhou, W.~B. Mori, and T.~Katsouleas.
\newblock {\em Physical Review Letters}, 96(16):165002, 2006.

\bibitem{lu2006pop}
W.~Lu, C.~Huang, M.~Zhou, M.~Tzoufras, F.~S. Tsung, W.~B. Mori, and
  T.~Katsouleas.
\newblock {\em Physics of Plasmas}, 13(5):056709, 2006.

\bibitem{rosenzweig1991}
J.~B. Rosenzweig, B.~Breizman, T.~Katsouleas, and J.~J. Su.
\newblock {\em Physical Review A}, 44(10), 1991.

\bibitem{plateau2012}
G.~R. Plateau, C.~G.~R. Geddes, D.~B. Thorn, M.~Chen, C.~Benedetti, E.~Esarey,
  A.~J. Gonsalves, N.~H. Matlis, K.~Nakamura, C.~B. Schroeder, S.~Shiraishi,
  T.~Sokollik, J.~van Tilborg, Cs~Toth, S.~Trotsenko, T.~S. Kim, M.~Battaglia,
  Th~Stoehlker, and W.~P. Leemans.
\newblock {\em Physical Review Letters}, 109(6):064802, 2012.

\bibitem{wiggins2010}
S.~M. Wiggins, R.~C. Issac, G.~H. Welsh, E.~Brunetti, R.~P. Shanks, M.~P.
  Anania, S.~Cipiccia, G.~G. Manahan, C.~Aniculaesei, B.~Ersfeld, M.~R. Islam,
  R.~T.~L. Burgess, G.~Vieux, W.~A. Gillespie, A.~M. MacLeod, S.~B. van~der
  Geer, M.~J. de~Loos, and D.~A. Jaroszynski.
\newblock {\em Plasma Physics and Controlled Fusion}, 52(12):124032, 2010.

\bibitem{umstadter1996}
D.~Umstadter, J.~K. Kim, and E.~Dodd.
\newblock {\em Physical Review Letters}, 76(12):2073, 1996.

\bibitem{vieira2011}
J.~Vieira, S.~F. Martins, V.~B. Pathak, R.~A. Fonseca, W.~B. Mori, and L.~O.
  Silva.
\newblock {\em Physical Review Letters}, 106(22):225001, 2011.

\bibitem{esarey1997}
E.~Esarey, R.~F. Hubbard, W.~P. Leemans, A.~Ting, and P.~Sprangle.
\newblock {\em Physical Review Letters}, 79(14):2682, 1997.

\bibitem{faure2006}
J.~Faure, C.~Rechatin, A.~Norlin, A.~Lifschitz, Y.~Glinec, and V.~Malka.
\newblock {\em Nature}, 444(7120):05393, 2006.

\bibitem{suk2001}
H.~Suk, N.~Barov, J.~B. Rosenzweig, and E.~Esarey.
\newblock {\em Physical Review Letters}, 86(6):1011, 2001.

\bibitem{geddes2008}
C.~G.~R. Geddes, K.~Nakamura, G.~R. Plateau, Cs~Toth, E.~Cormier-Michel,
  E.~Esarey, C.~B. Schroeder, J.~R. Cary, and W.~P. Leemans.
\newblock {\em Physical Review Letters}, 100(21):215004, 2008.

\bibitem{oz2007}
E.~Oz, S.~Deng, T.~Katsouleas, P.~Muggli, C.~D. Barnes, I.~Blumenfeld, F.~J.
  Decker, P.~Emma, M.~J. Hogan, R.~Ischebeck, R.~H. Iverson, N.~Kirby,
  P.~Krejcik, C.~O'Connell, R.~H. Siemann, D.~Walz, D.~Auerbach, C.~E. Clayton,
  C.~Huang, D.~K. Johnson, C.~Joshi, W.~Lu, K.~A. Marsh, W.~B. Mori, and
  M.~Zhou.
\newblock {\em Physical Review Letters}, 98(8):084801, 2007.

\bibitem{rowlands2008}
T.~P. Rowlands-Rees, C.~Kamperidis, S.~Kneip, A.~J. Gonsalves, S.~P.~D.
  Mangles, J.~G. Gallacher, E.~Brunetti, T.~Ibbotson, C.~D. Murphy, P.~S.
  Foster, M.~J.~V. Streeter, F.~Budde, P.~A. Norreys, D.~A. Jaroszynski,
  K.~Krushelnick, Z.~Najmudin, and S.~M. Hooker.
\newblock {\em Physical Review Letters}, 100(10):105005, 2008.

\bibitem{pak2010}
A.~Pak, K.~A. Marsh, S.~F. Martins, W.~Lu, W.~B. Mori, and C.~Joshi.
\newblock {\em Physical Review Letters}, 104(2):025003, 2010.

\bibitem{mcguffey2010}
C.~McGuffey, A.~G.~R. Thomas, W.~Schumaker, T.~Matsuoka, V.~Chvykov, F.~J.
  Dollar, G.~Kalintchenko, V.~Yanovsky, A.~Maksimchuk, K.~Krushelnick, V.~Yu
  Bychenkov, I.~V. Glazyrin, and A.~V. Karpeev.
\newblock {\em Physical Review Letters}, 104(2):025004, 2010.

\bibitem{liu2011}
J.~S. Liu, C.~Q. Xia, W.~T. Wang, H.~Y. Lu, Ch~Wang, A.~H. Deng, W.~T. Li,
  H.~Zhang, X.~Y. Liang, Y.~X. Leng, X.~M. Lu, C.~Wang, J.~Z. Wang,
  K.~Nakajima, R.~X. Li, and Z.~Z. Xu.
\newblock {\em Physical Review Letters}, 107(3):035001, 2011.

\bibitem{pollock2011}
B.~B. Pollock, C.~E. Clayton, J.~E. Ralph, F.~Albert, A.~Davidson, L.~Divol,
  C.~Filip, S.~H. Glenzer, K.~Herpoldt, W.~Lu, K.~A. Marsh, J.~Meinecke, W.~B.
  Mori, A.~Pak, T.~C. Rensink, J.~S. Ross, J.~Shaw, G.~R. Tynan, C.~Joshi, and
  D.~H. Froula.
\newblock {\em Physical Review Letters}, 107(4):045001, 2011.

\bibitem{hidding2012}
B.~Hidding, G.~Pretzler, J.~B. Rosenzweig, T.~Koenigstein, D.~Schiller, and
  D.~L. Bruhwiler.
\newblock {\em Physical Review Letters}, 108(3):035001, 2012.

\bibitem{emma2010}
P.~Emma, R.~Akre, J.~Arthur, R.~Bionta, C.~Bostedt, J.~Bozek, A.~Brachmann,
  P.~Bucksbaum, R.~Coffee, F.~J. Decker, Y.~Ding, D.~Dowell, S.~Edstrom,
  A.~Fisher, J.~Frisch, S.~Gilevich, J.~Hastings, G.~Hays, Ph~Hering, Z.~Huang,
  R.~Iverson, H.~Loos, M.~Messerschmidt, A.~Miahnahri, S.~Moeller, H.~D. Nuhn,
  G.~Pile, D.~Ratner, J.~Rzepiela, D.~Schultz, T.~Smith, P.~Stefan,
  H.~Tompkins, J.~Turner, J.~Welch, W.~White, J.~Wu, G.~Yocky, and J.~Galayda.
\newblock {\em Nature Photonics}, 4(9):641, 2010.

\bibitem{xu2014}
X.~L. Xu, J.~F. Hua, F.~Li, C.~J. Zhang, L.~X. Yan, Y.~C. Du, W.~H. Huang,
  H.~B. Chen, C.~X. Tang, W.~Lu, P.~Yu, W.~An, C.~Joshi, and W.~B. Mori.
\newblock {\em Physical Review Letters}, 112(3):035003, 2014.

\bibitem{lifei2013}
F.~Li, J.~F. Hua, X.~L. Xu, C.~J. Zhang, L.~X. Yan, Y.~C. Du, W.~H. Huang,
  H.~B. Chen, C.~X. Tang, W.~Lu, C.~Joshi, W.~B. Mori, and Y.~Q. Gu.
\newblock {\em Physical Review Letters}, 111(1):015003, 2013.

\bibitem{fonseca2002}
R.~A. Fonseca et~al.
\newblock volume 2331, pages 342--351.
\newblock Springer Berlin Heidelberg, 2002.

\bibitem{ammosov1986}
M.~V. Ammosov, N.~B. Delone, and V.~P. Krainov.
\newblock {\em Soviet Physics - JETP}, 64(6), 1986.

\bibitem{wang2002}
S.~Q. Wang, C.~E. Clayton, B.~E. Blue, E.~S. Dodd, K.~A. Marsh, W.~B. Mori,
  C.~Joshi, S.~Lee, P.~Muggli, T.~Katsouleas, F.~J. Decker, M.~J. Hogan, R.~H.
  Iverson, P.~Raimondi, D.~Walz, R.~Siemann, and R.~Assmann.
\newblock {\em Physical Review Letters}, 88(13):135004, 2002.

\end{thebibliography}
\end{document}